\documentclass[journal]{IEEEtran}

\usepackage{amsmath}%
\usepackage{amsfonts}%
\usepackage{amssymb}%
\usepackage{graphicx}
\usepackage{color}
\usepackage[usenames,dvipsnames]{pstricks}
\usepackage{dsfont,latexsym}
\usepackage{psfrag}
\usepackage[noadjust]{cite}
\usepackage{algorithmic}
\usepackage{algorithm}
\usepackage[colorlinks=true,linkcolor=black, citecolor=black,urlcolor=blue]{hyperref}
\usepackage{todonotes}

\begin{document}

		\title{Across-domains transferability of Deep-RED in de-noising and compressive sensing recovery of seismic data}

		\author{\IEEEauthorblockN{Nasser Kazemi}\\

\thanks{N. Kazemi is with the Schulich School of  Engineering, University of Calgary. Email: nasser.kazeminojadeh@ucalgary.ca.}}

	\maketitle

\begin{abstract}
In the past decade, deep learning algorithms gained a remarkable interest in the signal processing community. The availability of big datasets and advanced computational resources resulted in developing efficient algorithms. However, such algorithms are biased towards the training dataset. Thus, the transferability of deep-learning-based operators are challenging, especially when the goal is to apply the learned operator on a new dataset/domain. Lack of transferability of learned operator across domains hinders the applicability of deep learning algorithms in processing seismic data. Unlike camera images, the comprehensively labeled seismic datasets are not available. Moreover, from one task to another, the training parameters should be tuned. To remedy this shortcoming, we have developed a workflow that transfers the learned operator from the camera images to the seismic domain, without modifying its training parameters. The similarities in the algorithms and optimization methods in camera and seismic data processing allow us to do so. Accordingly, by incorporating feed-forward de-noising convolutional neural networks (DnCNN) in regularization by de-noising regularizer, we formulate two transferable optimization problems for de-noising and compressive sensing recovery of seismic data. Simulated and real-world data examples show the efficiency of our proposed workflow.

\end{abstract}
\renewcommand\IEEEkeywordsname{Index Terms}
\section{Introduction}
A wide variety of signal processing applications can be cast as a generic optimization problem with the cost function of 
\begin{equation}\label{n:1}
{\hat{\bf s}}=\underset{{\bf s}}{\operatorname{argmin}} \quad{f({\bf s})+g({\bf s})},
\end{equation}
where $f({\bf s})$ is a convex and differentiable function, and $g({\bf s})$ is a convex but possibly non-smooth function. 
For example, this cost function is encountered in constrained least-squares
problems~\cite{35,lines1984review,chen2015random}, sparse regularization problems~\cite{10,30,candes2005decoding,kazemi2014sparse,metzler2016denoising,vera2016compressive},
Fourier regularization problems~\cite{46,liu2004minimum}, 
alternating projection problems~\cite{38,beasley20123d},
split feasibility problems~\cite{13,15}, and total variation problems~\cite{19, 62,gholami2013fast}.

Recently, deep learning algorithms gained a remarkable interest in the signal processing community~\cite{krizhevsky2012imagenet,jia2017can,metzler2017learned,metzler2018prdeep,yu2019deep,pham2019automatic,li2019deep,khobahi2019deep,park2020automatic}. Deep learning algorithms approximate a complicated set of operators through building network architecture and model learning. Traditional machine/deep learning algorithms can learn and model tasks within a specific domain.  However, the learned task is not transferable to other domains. In other words, to use the learned network on a new dataset, the learning process should start from scratch, which results in wasting computational resources.

The generalizability issue of deep-learned tasks motivated machine learning community to start thinking about the transfer learning process across domains and tasks~\cite{mihalkova2007mapping,pan2009survey,yosinski2014transferable,tzeng2015simultaneous}. Transfer learning is usually performed within the same domain whenever there are not sufficient training data for modeling the complexities of the underlying system~\cite{siahkoohi2019importance,park2020automatic}. However, the question that we are interested in is the transferability of learned tasks across domains. A feature that makes this a possibility is the similarity in the structures of the two related domains~\cite{mihalkova2007mapping,domingos2007structured}. For example, the two tasks in the relational domains can be modeled with the same formula. The similarities in signal processing formula on data with different natures, e.g., seismic data, camera images, and MRI images, is a key observation that could help pave the way for generalizing the learned operators across domains. As we discussed before, Equation~\eqref{n:1} is a general formula that we routinely use to recover the signal of interest regardless of the nature of the acquisition system and data. Accordingly, if machine-learned operators are incorporated within the same framework, then there is a strong possibility that the learned operator can be transferred across domains without or with little modifications. In this paper, we show the transferability of feed-forward de-noising convolutional neural networks (DnCNN) learned operator, proposed by~\cite{zhang2017beyond},  across camera image and seismic data processing domains.The transferability of DnCNN learned noise estimating operators, which are trained by using camera images only, is exemplified by showing its applications on de-noising and compressive sensing recovery of seismic data. 
 
  The paper is organized as follows. We start the paper by introducing the general optimization problem of interest. Then, we explain a class of regularization techniques, called regularization by de-noising (RED)~\cite{romano2017little}, which is suitable for modern signal processing. Next, we discuss the DnCNN learning framework and incorporate the deep-learning-based operator into RED regularizer. Moreover, two specific formulations of the general optimization problem, represented in Equation~\eqref{n:1}, are developed for de-noising and compressive sensing recovery of seismic data. The algorithms explore the potential of across-domains transferability of deep-learned RED from camera image processing to seismic signal processing.
  Finally, we present the concluding remarks.  
 \section{general problem statement}
 Bayesian estimation of desired output ${\bf s}$ given the noisy measurements ${\bf m}$ can be achieved by using a posterior conditional probability, $P({\bf s}|{\bf m})$. We usually use the maximum a posteriori probability estimator. Following Bayes' rule and assuming that $P({\bf m})$ is not a function of ${\bf s}$, the task of estimating the desired output is turned into an optimization
  \begin{equation}\label{n:02}
  {\hat{\bf s}}=\underset{{\bf s}}{\operatorname{argmax}} \;P({\bf s}|{\bf m})P({\bf s})=\underset{{\bf s}}{\operatorname{argmin}} \; -log\{P({\bf s}|{\bf m})\}-log\{P({\bf s})\}.
\end{equation}  
 where $-log\{P({\bf s}|{\bf m})\}$ is log-likelihood term, and $-log\{P({\bf s})\}$ is $\it{prior}$ or regularization term. Note that $-log$ function is also assumed to be monotonically decreasing. Comparing Equations~\eqref{n:1} and~\eqref{n:02} shows that 
\begin{equation}\label{n:03}
  g({\bf s})=\lambda{\cal R}({\bf s})=-log\{P({\bf s})\},
\end{equation}   
 where $\lambda$ is a regularization parameter, and ${\cal R}({\bf s})$ is a regularization term. Hence, the general optimization problem in Equation~\eqref{n:1} can be re-written as
\begin{equation}\label{n:04}
{\hat{\bf s}}=\underset{{\bf s}}{\operatorname{argmin}} \quad{f({\bf s})+\lambda{\cal R}({\bf s})}.
\end{equation}  
A large class of signal processing optimization problems can be solved by choosing proper functions for $f({\bf s})$, and ${\cal R}({\bf s})$. Recall that we assumed the $f({\bf s})$ function is convex and differentiable and ${\cal R}({\bf s})$ is convex but possibly non-smooth. Given the mentioned assumptions, this general cost function can be efficiently solved by using the proximal forward-backward splitting (FBS) algorithm~\cite{combettes2005signal}. This method is also known as the proximal gradient. The non-smoothness property of the regularization term stops us from using a classical gradient descent method. However, for a large class of non-differentiable but convex functions, there is a proximal operator such that
\begin{equation}\label{n:05}
prox_ g({\bf s},{\bf z})=\underset{{\bf s}}{\operatorname{argmin}} \quad{\tau g({\bf s})+\frac{1}{2}\|{\bf s}-{\bf z}\|_2^2},
\end{equation}      
where $\tau$ is step-size, and ${\bf z}$ is a backward gradient-descent step. To solve the proximal operator, we only need to calculate the sub-gradient (generalized gradient) of $g$. In cases where the proximal operator can be evaluated, the FBS algorithm can efficiently solve the general cost function represented in Equation~\eqref{n:04}. The details of the FBS algorithm are presented in Algorithm~\ref{alg1}. 
\begin{algorithm}[b]{
 \caption{FBS algorithm for solving Equation~\eqref{n:04}}
  \label{alg1}
 \begin{algorithmic}[1]
 \REQUIRE $f(\cdot)$, ${\cal R}(\cdot)$, $\lambda$, ${\bf s}^0$, $k=0$\\
 \mbox {{\bf While not converged}}\\
\STATE ${\hat{\bf s}}^{k+1}={\bf s}^k-\tau^k\nabla f({\bf s}^k)$\\
 \STATE ${\bf s}^{k+1}=prox_g({\hat{\bf s}}^{k+1},\tau^k)=\underset{{\bf s}}{\operatorname{argmin}} \quad{\tau^k \lambda {\cal R}({\bf s})+\frac{1}{2}\|{\bf s}-{\hat{\bf s}}^{k+1}\|_2^2}$\\
\STATE \textit{Update} $k\leftarrow k+1$ \\
  \mbox{{\bf If converged}}\\
${\bf s}\leftarrow {{\bf s}}^k$\\
 \end{algorithmic} }
 \end{algorithm}
 In the next section, we introduce a generalized de-nosing regularization function for ${\cal R}({\bf s})$ that can be incorporated into Equation~\eqref{n:04} to solve different signal processing problems.
       
 \section{Regularization by De-noising}  
Two observations guide us toward defining a general and efficient de-noising regularizer. 
Let's assume that the clean input, e.g., image, lives on a manifold ${\cal M}$. Then, the first observation is that adding noise to the clean image moves the image, with high probability, out of the manifold along the direction normal to ${\cal M}$~\cite{romano2017little}. Hence, the ideal de-noising operator projects the noisy image back into its manifold, leaving the noise component in the normal direction to the manifold. In other words, at the end of the day, the noise component is orthogonal to the clean image. The second observation is that most de-noising operators can be modeled as the action of input-dependent pseudo-linear operator ${\bf W}(\cdot)$ on the input~\cite{milanfar2012tour} 
\begin{equation}\label{n:1-1}
{\cal D}({\bf m})={\bf W}({\bf m}){\bf m}.
\end{equation} 
where ${\cal D}({\cdot})$ is a de-noising operator. Equation~\eqref{n:1-1} is valid for most de-noising methods~\cite{romano2017little}. By combining these two observations, we develop an input-adaptive Laplacian regularizer, which is the extension of classical Laplacian smoothness regularizer~\cite{lagendijk2012iterative}
\begin{equation}\label{n:1-2}
{\cal R}({\bf s})={\bf s}^T{\bf L}({\bf s}){\bf s}={\bf s}^T({\bf I}-{\bf W}({\bf s})){\bf s}={\bf s}^T({\bf s}-{\cal D}({\bf s})),
\end{equation} 
where ${\bf I}$ is an identity matrix, and $T$ stands for transpose. The regularizer ${\cal R}({\bf s})$, in Equation~\eqref{n:1-2}, is called regularization by de-noising (RED)~\cite{romano2017little}. Note that the RED function promotes orthogonality between the predicted noise and the input. 
The definition of a de-noising function ${\cal D}({\cdot})$ can also be extended to any random functions as long as they obey homogeneity and passivity conditions~\cite{romano2017little}. This relaxation allows us to incorporate powerful de-noising operators into the RED regularizer. Recent works show the benefits of RED regularizer in deep-learning-based image processing applications~\cite{metzler2018prdeep,mataev2019deepred}. The inclusion of deep-learning-based de-noising operators within RED has extensive potential that needs to be explored. In the next section, we introduce a deep-learning-based de-noising operator that can be incorporated in RED.         
 \section{Feed-forward de-noising convolutional neural networks}
\begin{figure}[h]%
 \vspace{-0cm}
\centering  
   \includegraphics[width=.5\textwidth]{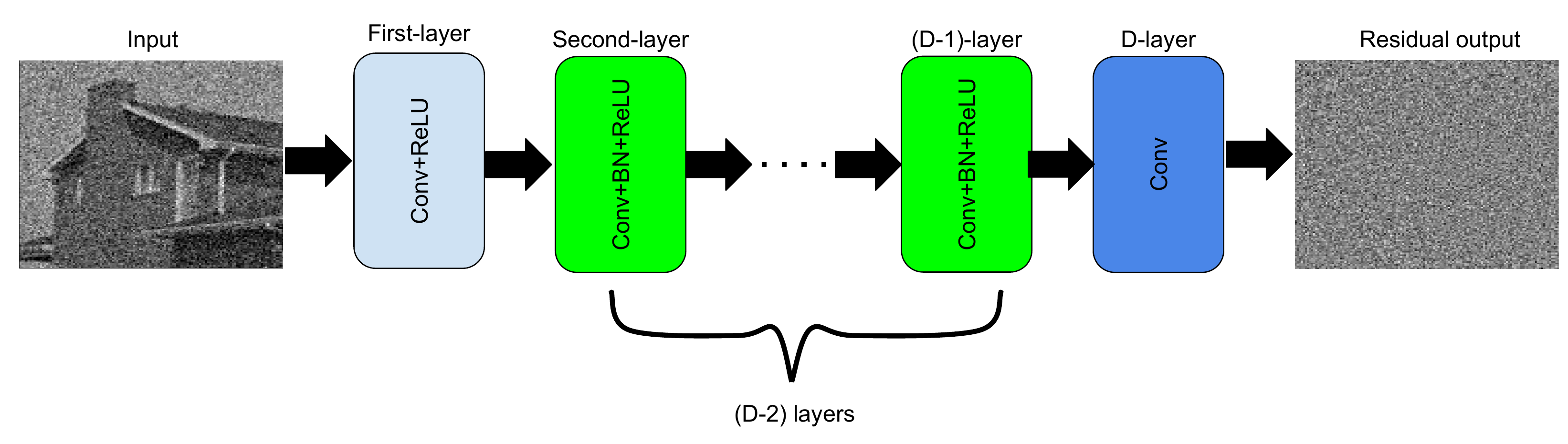}
   \vspace{-.75cm}  
   \caption{The schematic representation of DnCNN architecture.}
\label{fig1}%
\end{figure}

 Feed-forward de-noising convolutional neural networks (DnCNNs) is a residual learning~\cite{he2016deep} method, which is combined with batch normalization to increase performance and learning speed~\cite{zhang2017beyond}.  In other words, the network learns to remove the latent clean image, which is hidden in the layers of the network, and predict the noise component as a residual output. DnCNN can be used as a blind Gaussian denoiser, meaning it does not require to be learned on data with known noise levels. Also, the network has a simple structure and can be efficiently parallelized to take advantage of advanced computational resources. Hence, we adopt the DnCNN method as a noise estimating operator. DnCNN works as follows. Consider a noisy gray-colored image, i.e., it has only one image channel,   
 \begin{equation}\label{n:2}
 {\bf m}={\bf s}+{\bf n},
 \end{equation}
 where ${\bf m}$ is a noisy image, ${\bf s}$ is a clean image, and ${\bf n}$ is an additive white Gaussian noise. In this model, DnCNN acts as a noise estimating operator
 \begin{equation}\label{n:3}
 {\cal L}({\bf m})\approx{\bf n},\quad \mbox{and}\quad {\bf s}\approx {\bf m}-{\cal L}({\bf m}),
 \end{equation}
 where ${\cal L}(\cdot)$ is the DnCNN-learned mapping operator. To learn the model, DnCNN minimizes the averaged mean-squared-error between the estimated and ground-truth noise components by solving
 \begin{equation}\label{n:4}
  {\hat{\bf \Theta}}=\underset{{\bf \Theta}}{\operatorname{argmin}} \quad \frac{1}{2N}\;\sum_{j=1}^{N}{\| {\cal L}({\bf m}_j;{\bf \Theta})-({\bf m}_j-{\bf s}_j)\|_F^2},
\end{equation}
where ${\bf \Theta}$ are the trainable parameters in DnCNN, ${\bf m}_j$ and ${\bf s}_j$ are the $j^{th}$ noisy-clean training image (patch) pairs, $N$ is the total number of images in the training library, and $F$ is Frobenius norm. 
 Figure~\ref{fig1} shows the schematics of the DnCNN architecture for learning the mapping function. 
 
 The DnCNN has a deep architecture with three types of layers. The layers are built by combining convolutional layer (Conv) with Rectifier Linear Unit (ReLU)~\cite{krizhevsky2012imagenet}, and batch normalization (BN)~\cite{ioffe2015batch}. The first layer is Conv+ReLU with 64 filters of size $3\times3\times1$,  which generates 64 feature maps, and ReLU is used to promote non-linearity. ReLU acts as a function that outputs the positive values of the input and zeros out the negative part, i.e., ReLU($input$)=$max(0,input)$. From the second layer to the $D-1$ layer, we have Conv+ BN+ReLU, where $D$ is the depth of DnCNN architecture. In these layers, we have 64 filters of size $3\times3\times64$, and then batch normalization and ReLU functions are applied to the filters. Finally, the last layer is the Conv layer with a single filter of size $3\times3\times64$ which reconstructs the output. Note that the gray images are used for training purposes. In a nutshell, DnCNN is a combination of residual learning formulation and batch normalization. To learn the training parameters, the optimization problem represented in Equation~\eqref{n:4} is solved with Adam algorithm~\cite{kingma2014adam}. In the next section, we incorporate the DnCNN mapping function in RED and use it for de-nosing seismic data. 
\section{Seismic de-noising}
Suppressing noise in seismic recordings is an active area of research~\cite{canales1984random,abma1995lateral,trickett2008f,oropeza2011simultaneous,naghizadeh2012multicomponent,chen2015random,yu2019deep}. In seismic recordings with several channels, the signal shows some form of spatial coherency, and the noise component can be either spatially coherent or random. 
The de-noising methods can be local or non-local~\cite{bonar2012denoising,chen2015random}. Several algorithms take advantage of the transformation techniques to separate the noise from the signal in the transformed domain and apply de-noising by filtering or inversion~\cite{canales1984random,abma1995lateral,trickett2008f,oropeza2011simultaneous,naghizadeh2012multicomponent}. These algorithms can be used on different datasets with different natures. 

To give an idea about the similarity of the algorithms across domains, it suffices to mention that non-local means algorithm is used to de-noise camera images, MRI images, radar data, microscopy data, and seismic data~\cite{buades2005review,coupe2008optimized,buades2010image,wei2010optimized,deledalle2010nl,huang2011sparse,bonar2012denoising}. However, despite the huge progress in developing such algorithms, they require several assumptions that may not be valid from one dataset to another. Recently, authors also explored the benefits of using deep-learning-based de-noising algorithms~\cite{yu2019deep,zhu2019seismic,richardson2019seismic}. However, these algorithms can only deal with specific datasets that were in the training set, and also they are not blind to noise level, meaning they cannot handle datasets with variable signal-to-noise ratio (SNR).  

To take the best of two worlds, i.e, conventional de-noising and deep-learning-based de-noising, we define a specific form of the general optimization problem of Equation~\eqref{n:04} for random noise suppression in seismic data. We use the RED regularizer to promote orthogonality between the noise component and the clean data and incorporate DnCNN learned operator as a noise estimator within the RED. This regularizer is called Deep-RED. Also, $\ell_2$ norm is used as a misfit function between the clean and noisy datasets. Hence, by using the signal-noise model represented in Equation~\eqref{n:2} and the general cost function represented in Equation~\eqref{n:04}, we propose to solve
\begin{equation}\label{n:5}
{\hat{\bf s}}=\underset{{\bf s}}{\operatorname{argmin}} \quad{\|{\bf m}-{\bf s}\|_2^2+\lambda{\bf s}^T({\bf s}-{\cal D}({\bf s}))},
\end{equation}          
 where ${\cal D}({\bf s})={\bf s}-{\cal L}({\bf s})$. Equation~\eqref{n:5} can be solved with Algorithm~\ref{alg1} by setting $f({\bf s})=\|{\bf m}-{\bf s}\|_2^2$ and ${\cal R}({\bf s})={\bf s}^T({\bf s}-{\cal D}({\bf s}))$. Note that the regularization term is minimum when the signal and noise components are orthogonal to each other or when the data is clean, i.e., ${\cal D}({\bf s})={\bf s}$. It is worth mentioning that the ${\cal L}(\cdot)$ is the DnCNN based operator for noise estimation. The ${\cal L}(\cdot)$ operator is merely learned on camera images. Accordingly, by incorporating ${\cal L}(\cdot)$ operator in RED and using the FBS algorithm to solve the optimization problem in Equation~\eqref{n:5}, we are seeking to evaluate the performance of DnCNN across camera image processing and seismic signal processing domains. The misfit function of Equation~\eqref{n:5} can also be evaluated in the transformed domain, where the noise and signal are further separated 
 \begin{equation}\label{n:6}
{\hat{\bf s}}=\underset{{\bf s}}{\operatorname{argmin}} \quad{\|{\bf y}-{\bf A}{\bf s}\|_2^2+\lambda{\bf s}^T({\bf s}-{\cal D}({\bf s}))},
\end{equation}         
 where ${\cal D}({\bf s})={\bf s}-{\cal L}({\bf s})$, ${\bf y}={\bf A}{\bf m}$, and ${\bf A}$ is a transformation matrix.  
In this case, Equation~\eqref{n:6} can be also solved with FBS algorithm by setting $f({\bf s})=\|{\bf y}-{\bf A}{\bf s}\|_2^2$, and ${\cal R}({\bf s})={\bf s}^T({\bf s}-{\cal D}({\bf s}))$. Note that to solve the proximal cost function in the second step of the FBS algorithm, we require to calculate the derivative of the cost function with respect to ${\bf s}$. The gradient of the second term is straightforward, but the gradient of ${\cal R}({\bf s})$  with DnCNN can be challenging. Fortunately, Romano et al., show that if the denoiser satisfies the local homogeneity condition, we get $\nabla_{{\bf s}}{\cal D}({\bf s}){\bf s}={\cal D}({\bf s})$~\cite{romano2017little}. In other words, we have $\nabla_{{\bf s}}{\cal R}({\bf s})=2({\bf s}-{\cal D}({\bf s}))$. There is also the possibility of using Monte Carlo based algorithms to approximate the derivate of the de-noising operator~\cite{ramani2008monte,metzler2016denoising}. The details of the algorithm for solving Equation~\eqref{n:6} are presented in Algorithm~\ref{alg2}. There are fast solvers such as FASTA that can provide efficient solutions for the FBS algorithm~\cite{goldstein2014field}. For a detailed analysis of the convergence property of the algorithm and its connection with the choice of step-size, interested readers are referred to~\cite{goldstein2014field}. In the next section, we show the application of Deep-RED in compressed sensing recovery of seismic data. 
 \begin{algorithm}{
 \caption{FBS algorithm for solving Equation~\eqref{n:6}}
  \label{alg2}
 \begin{algorithmic}[1]
 \REQUIRE ${\bf y}$, ${\cal D}(\cdot)$, ${\bf A}$, $\lambda$, ${\bf s}^0$, $\tau^0$, $k=0$\\
 \mbox {{\bf While not converged}}\\
\STATE ${\hat{\bf s}}^{k+1}={\bf s}^k-\tau^k\;{\bf A}^*({\bf A}{\bf s}^k-{\bf y})$\\
 \STATE ${\bf s}^{k+1}=\underset{{\bf s}}{\operatorname{argmin}} \quad{\tau^k \lambda \;{\bf s}^T({\bf s}-{\cal D}({\bf s}))+\frac{1}{2}\|{\bf s}-{\hat{\bf s}}^{k+1}\|_2^2}$\\
\STATE \textit{Update} $\tau^{k+1} \textit{with line search method}$ \\  
\STATE \textit{Update} $k\leftarrow k+1$ \\
  \mbox{{\bf If converged}}\\
${\bf s}\leftarrow {{\bf s}}^k$\\
 \end{algorithmic} }
 \end{algorithm} 
\section{Compressive sensing recovery of seismic data} 
 Compressive sensing is a technique that recovers the signal of interest ${\bf s}_{q\times 1}$ from its compressedly sampled measurements ${\bf m}_{p\times 1}$ under a measurement matrix ${\bf \Phi}_{p\times q}$
  \begin{equation}\label{n:7}
 {\bf m}={\bf \Phi}{\bf s}+{\bf n},  
 \end{equation}
 where $p \ll q$. In some cases, the signal can be sparsely represented in a transformed domain. Then, Equation~\eqref{n:7} can be written as 
  \begin{equation}\label{n:8}
 {\bf m}={\bf \Phi}{\bf \Psi}{\bf x}+{\bf n},  
 \end{equation} 
 where ${\bf s}={\bf \Psi}{\bf x}$, ${\bf \Psi}$ is inverse transformation matrix, and ${\bf x}$ is a sparse representation of signal of interest in the transformed domain. To recover the signal, sparsity-based compressive sensing algorithms solve the following optimization problem
 \begin{equation}\label{n:9}
{\hat{\bf x}}=\underset{{\bf x}}{\operatorname{argmin}} \quad{\|{\bf m}-{\bf A}{\bf x}\|_2^2+\lambda\; \|{\bf x}\|_1},
\end{equation}    
where ${\bf A}={\bf \Phi}{\bf \Psi}$, and $\ell_1$ norm is used to promote sparsity on ${\bf x}$. This cost function is well known as basis pursuit de-noising or Lasso~\cite{donoho2006compressed}. The iterative approach to solve the cost function of Equation~\eqref{n:9} involves two steps
\begin{equation}\label{n:10}
\begin{split}
{\bf z}^k={\bf m}-{\bf A}{\bf x}^k\\
{\bf x}^{k+1}=\gamma_\tau({\bf A}^*{\bf z}^k+{\bf x}^k),
\end{split}
\end{equation}
where $\gamma_\tau$ is a thresholding operator, ${\bf z}^k$ is residual, $k$ is iteration number, and $*$ sands for conjugate transpose.
 Sparsity-based compressive sensing recovery of signals is deeply studied in the literature~\cite{donoho2006compressed,baraniuk2007compressive,ji2008bayesian,chartrand2008iteratively,dai2009subspace,eldar2012compressed}. 

In this paper, however, we are interested in across-domains transferability of the Deep-RED denoiser in compressive sensing recovery of seismic data. Hence, our ideal algorithm would be the one that uses Deep-RED denoiser in its formulation.
Accordingly, we assume that the data ${\bf s}$ is compressed under the compression matrix ${\bf A}_c$ 
\begin{equation}\label{n:11}
{\bf y}_c={\bf A}_c{\bf s}+{\bf n},
\end{equation}
where ${\bf y}_c$ is compressed measurements. The compression matrix ${\bf A}_c$ is a $p\times q$ matrix with $p\ll q$. We also assume that the noise component ${\bf n}$ is orthogonal to un-compressed signal ${\bf s}$. To recover the un-compressed signal, we propose to solve
 \begin{equation}\label{n:12}
{\hat{\bf s}}=\underset{{\bf s}}{\operatorname{argmin}} \quad{\|{\bf y}_c-{\bf A}_c{\bf s}\|_2^2+\lambda\; {\bf s}^T({\bf s}-{\cal D}({\bf s}))},
\end{equation}       
where ${\cal D}({\bf s})={\bf s}-{\cal L}({\bf s})$, and ${\cal L}(\cdot)$ is DnCNN operator. Equation~\eqref{n:12} can also be solved by Algorithm~\ref{alg2}, by replacing ${\bf A}_c\rightarrow{\bf A}$, ${\bf y}_c\rightarrow{\bf y}$, and setting ${\bf s}^0={\bf A}_c^*{\bf y}_c$.

 Note that in each iteration the updated solution may not have residual noise that satisfies the Gaussian distribution. To address this issue, Metzler et al., applied the approximate message passing~\cite{maleki2011approximate} concept, and added Onsager correction term into the compressive sensing recovery algorithm~\cite{metzler2016denoising}. The Onsager correction term guarantees the Gaussianity of the residual in each iteration. Nonetheless, here, we are not concerned with this shortcoming of the algorithm, and we merely focus on across-domains transferability of the Deep-RED regularization.  
\section{Numerical examples}
We start the section by analyzing the local homogeneity property of the DnCNN operator. The DnCNN architecture consists of $20$ layers, and its training parameters are optimized by using the Adam algorithm. The DnCNN operator is trained on gray-colored camera images\footnote{We use the pre-trained DnCNN operators which are trained on gray-colored camera images. The package is downloadable from~\url{https://github.com/ricedsp/prDeep/tree/master/Packages/DnCNN}. The package uses the DnCNN method developed by Zhang et al.,~\cite{zhang2017beyond}, and trains the DnCNN operators on data with selective ranges of standard deviations of noise. }. 
Then, we show the performances of the direct application of DnCNN operator and Deep-RED regularization, i.e., Algorithm~\ref{alg2}, in de-noising seismic data. In the de-noising application, we also show through Monte Carlo simulations the sensitivity of DnCNN operator and Deep-RED regularization to different noise realizations. Later, we use Algorithm~\ref{alg2} to recover the compressively sensed seismic data. The methods are tested on synthetic and real datasets.  
\subsection{Local homogeneity of DnCNN}
The local homogeneity of DnCNN operator, i.e., ${\cal L}(\cdot)$, is defined as
\begin{equation}\label{n:13}
{\cal L}({\bf s}+\epsilon {\bf s})\approx (1+\epsilon){\cal L}({\bf s}),
\end{equation}
where $\epsilon$ is a very small number. The metric that we use to evaluate the local homogeneity is 
\begin{equation}\label{n:14}
lh=\frac{\|{\cal L}({\bf s}+\epsilon {\bf s})-(1+\epsilon){\cal L}({\bf s})\|_2^2}{\|{\cal L}({\bf s})\|_2^2},
\end{equation}
where $lh$ is local homogeneity factor. The smaller $lh$ factor means that the operator satisfies the local homogeneity property, hence $\nabla_{{\bf s}}{\cal D}({\bf s}){\bf s}\approx{\cal D}({\bf s})$ is valid. This property allows us to efficiently use the Deep-RED regularizer in processing seismic data, where the gradient of the proximal operator is easy to compute. It is worth mentioning that in all of the numerical examples, we only use a single iteration to solve the proximal operator in step 2 of Algorithm~\ref{alg2}.  

\begin{figure}[h]%
 \vspace{-0.2cm}
\centering  
   \includegraphics[width=0.5\textwidth]{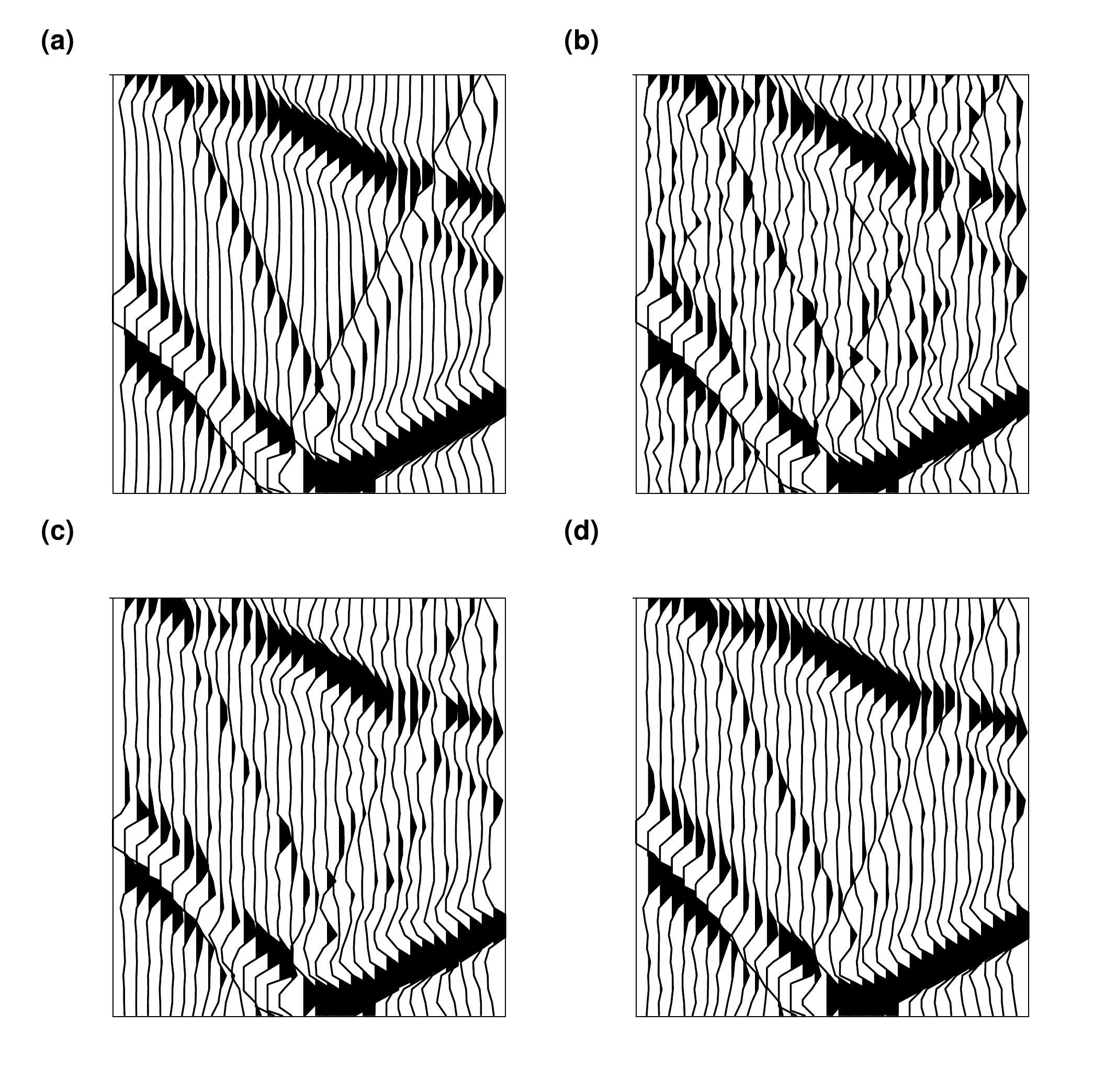}
   \vspace{-.75cm}  
   \caption{Performances of DnCNN and Deep-RED denoisers in de-noising seismic data. a) Clean data. b) Adjoint estimated data. c) DnCNN estimated data. d) Deep-RED estimated data. }
\label{fig2}%
\end{figure}

To check the local homogeneity property of DnCNN, we use Monte Carlo simulations and generate $100$ realizations of the added white Gaussian noise with $SNRs=1,2,3,4$. Similar to the work of~\cite{kazemi2016surface}, we define $SNR=\frac{a_{rms}^2}{\sigma_n^2}$, where $a_{rms}$ is the root-mean-square of the clean signal, and $\sigma_n^2$ is the standard deviation of noise. These noise realizations are added to the clean data shown in Figure~\ref{fig2}a. The seismic data has 32 channels and 32 time samples. Note that DnCNN has no assumption about the linearity of the events in the data. However, for simplicity, we model seismic data with linear events, which show random amplitude and slope. The DnCNN operators are trained on different ranges of standard deviations of noise in the camera images. The operators are trained on data with noise levels at intervals of $10$ in standard deviations (i.e., $\sigma_n^2=0-10, \sigma_n^2=10-20$, and so on).
We test these pre-trained DnCNN operators directly on the noise corrupted seismic data, and for each class of data with specific SNR, we calculate $lh$ factors by using Equation~\eqref{n:14} and report their mean value.
We choose $\epsilon=0.001$. The $lh$ factors for $SNRs=\{1,2,3,4\}$ are $lh=10^{-8}\times\{1.346, 1.059, 1.340, 1.416\}$, respectively. The small values of $lh$ factors show that the DnCNN operators have local homogeneity property. 
\vspace{-.2cm}   
\subsection{DnCNN and Deep-RED based de-noising}   
To analyze the performances of the de-noising methods, we introduce a new metric 
\begin{equation}\label{n:14}
Q=10 \log {\frac{\|{\bf s}^{true}\|_2^2}{\|{\bf s}^{est}+{\bf s}^{true}\|_2^2}},
\end{equation}
where ${\bf s}^{est}$ is estimated clean signal, ${\bf s}^{true}$ is ground-truth clean signal, and $Q$ is quality of reconstruction.

We evaluate the performances of the DnCNN operators, which are only trained on camera images, in estimating the noise in seismic data. To generate noisy measurements ${\bf y}_{p\times 1}$, we first transform the clean data ${\bf s}_{q\times 1}$, Figure~\ref{fig2}a, by using a i.i.d matrix ${\bf A}_{p\times q}$, where $\frac{p}{q}=8$. Note that ${\bf s}$ is the vectorized version of seismic section with $32$ channels and $32$ times samples, hence, $q=1024$. Later, we add white Gaussian noise to the transformed data to generate noisy measurements with $SNR=2$. We assume that the standard deviation of noise in unknown. Accordingly, DnCNN operators are independently applied to the data, and the operator that resulted in the highest quality of reconstruction is reported.  Figure~\ref{fig2}b shows the adjoint solution, i.e., ${\bf A}^T{\bf y}$. The adjoint solution is a good initial estimate for our de-noising algorithm (i.e., Algorithm~\ref{alg2}). Figure~\ref{fig2}c shows the direct application of the de-noising operator ${\cal D}(\cdot)$ on the adjoint image, i.e., ${\cal D}({\bf A}^T{\bf y})={\bf A}^T{\bf y}-{\cal L}({\bf A}^T{\bf y})$.  The de-noising operator estimates the clean data, however, the quality of reconstruction is poor ($Q=5.10 (dB)$). Later, we incorporate ${\cal D}(\cdot)$ operator within RED to de-noise the data (Figure~\ref{fig2}d). In this case, the quality of reconstruction is $Q=14.65(dB)$. In Algorithm~\ref{alg2}, we tune the regularization parameter $\lambda$ by trial and error, however, $\chi^2$ test or generalized cross-validation methods can provide the optimal regularization parameter, automatically. In all of the synthetic examples, we use $\lambda=0.01$. The higher quality of reconstruction by using Deep-RED regularizer shows that the DnCNN de-noising operator learned on camera images can be transferred to seismic data processing domain, efficiently. 
\begin{table}[t]
\renewcommand{\arraystretch}{1.3}
\caption{Sensitivity analysis of DnCNN and Deep-RED denoisers. }
\label{tab1}
\centering
\resizebox{9cm}{!} {
\begin{tabular}{lll}
    \hline
    $SNR$ & $Q_{DnCNN}(dB)$ & $Q_{Deep-RED}(dB)$\\
    \hline
    \hline
     $1$ &  $ 4.80 \pm 0.07 $  &  $8.72 \pm 0.21$ \\
    \hline
    $2$ &  $ 5.08 \pm 0.04 $  &  $14.61 \pm 0.20$  \\
   \hline
   $3$ &  $ 5.16 \pm 0.04 $  &  $18.13 \pm 0.19$  \\
   \hline
   $4$ &  $ 5.24 \pm 0.02 $  &  $20.60 \pm 0.20$  \\
   \hline
\end{tabular}
}
\end{table}
To better understand the stability and performances of Algorithm~\ref{alg2}, we use Monte Carlo simulations, similar to the approach implemented for estimating the local homogeneity property, and report mean and standard deviations of quality of reconstruction for each SNR. Table~\ref{tab1} summaries the results. The results show that by using Algorithm~\ref{alg2}, we improve the quality of reconstruction, dramatically. In each realization, we implement the best-trained operator and report the highest quality of reconstruction. Even though the results of the application of the DnCNN-based denoiser, directly on the data, clarify that the DnCNN operator is capable of estimating the noise. However, its performance is not satisfactory when compared to that of Algorithm~\ref{alg2}. This observation points us toward using more robust algorithms that take advantage of DnCNN in their formulation and avoid directly implementing them as a black box. Moreover, performances of Algorithm~\ref{alg2} can be improved, if we update the training parameters of the pre-trained DnCNN operators by using a limited number of labeled seismic data. Considering that the pre-trained operators can de-noise the seismic data, to some extent, there is no need for huge labeled seismic data or exhaustive training. In this paper, we simply focused on using the pre-trained DnCNN operators without modifying their training parameters.

\begin{figure}[]%
 \vspace{-0cm}
\centering  
   \includegraphics[width=.52\textwidth]{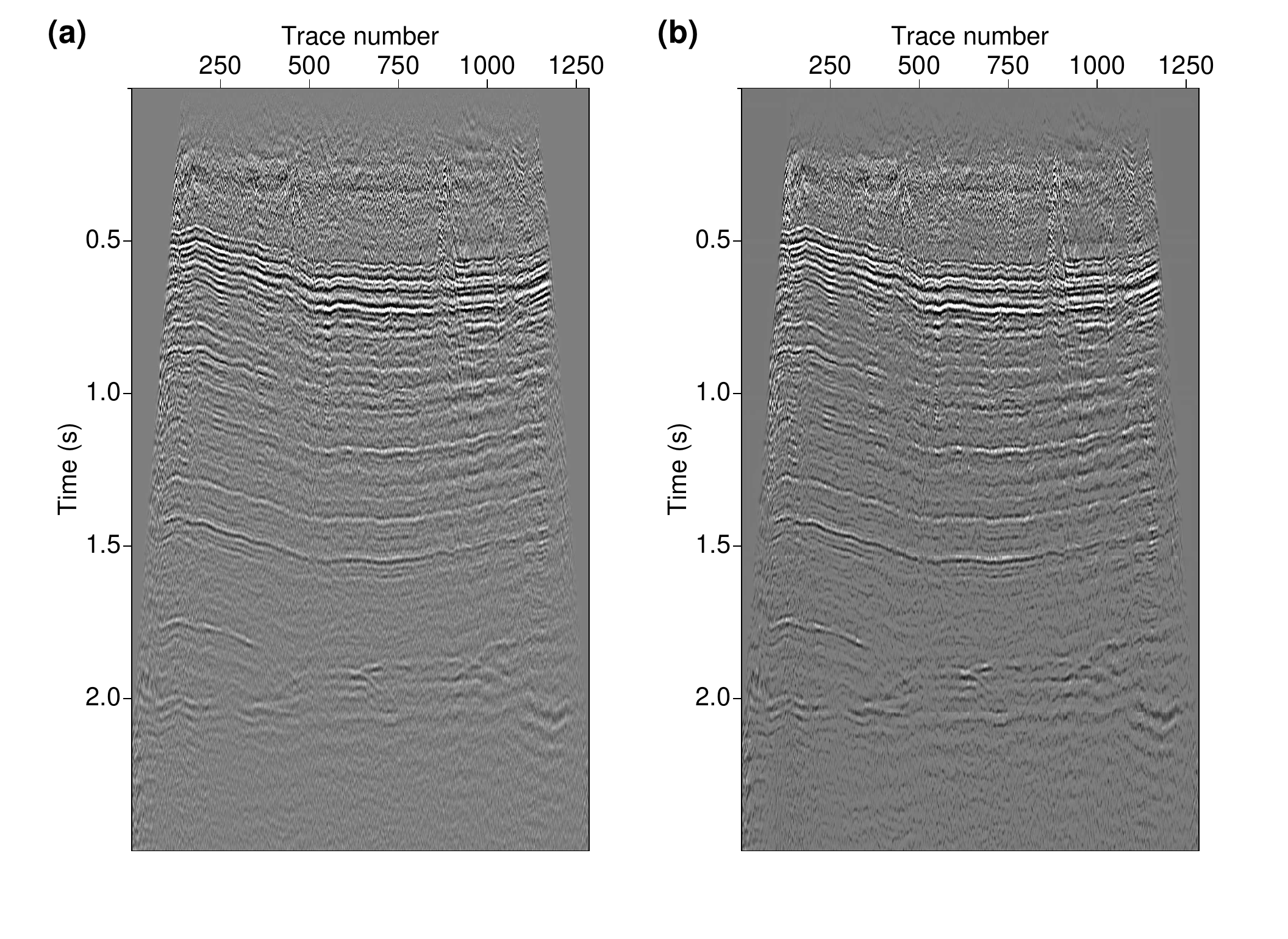}
   \vspace{-.75cm}  
   \caption{Performance of Deep-RED denoiser in de-noising real seismic data. a) Noisy data. b) De-noised data by using Deep-RED regularizer. }
\label{fig3}%
\end{figure}

The method is also successfully applied to real data. The real data shown in Figure~\ref{fig3}a is the stack section of processed land data provided by Geofizyka Torun Sp. Z.o.o, Poland~\footnote{Data is publicly available from~\url{http://www.freeusp.org/RaceCarWebsite/TechTransfer/Tutorials/Processing_2D}. We used Madagascar open-source software package~\cite{fomel2013madagascar}, which is freely available from~\url{http://www.ahay.org}, to process the data.}. Data has $751$ time samples with the sampling rate of $2(ms)$, and $1285$ channels. This data is also used in~\cite{liu2015signal} for de-noising purposes. We apply Algorithm~\ref{alg2} on patches with $128$ channels and $128$ time samples, without overlapping (bottom and left-corner patches are smaller). In each patch, we use all of the pre-trained DnCNN operators individually. In the real data, we do not know the ground-truth signal, hence, we use the solution that results in a maximum reduction in the cost function. All of the parameters, except the DnCNN operators are kept the same. Here, we use $\lambda=0.5$, and $\frac{p}{q}=4$. Figure~\ref{fig3}b shows the de-noising result of Deep-RED by using Algorithm~\ref{alg2}. The method suppresses the random noise component in the data, efficiently. 
\subsection{Deep-RED based compressive sensing recovery}
\begin{figure}[t]%
 \vspace{-0cm}
\centering  
   \includegraphics[width=0.5\textwidth]{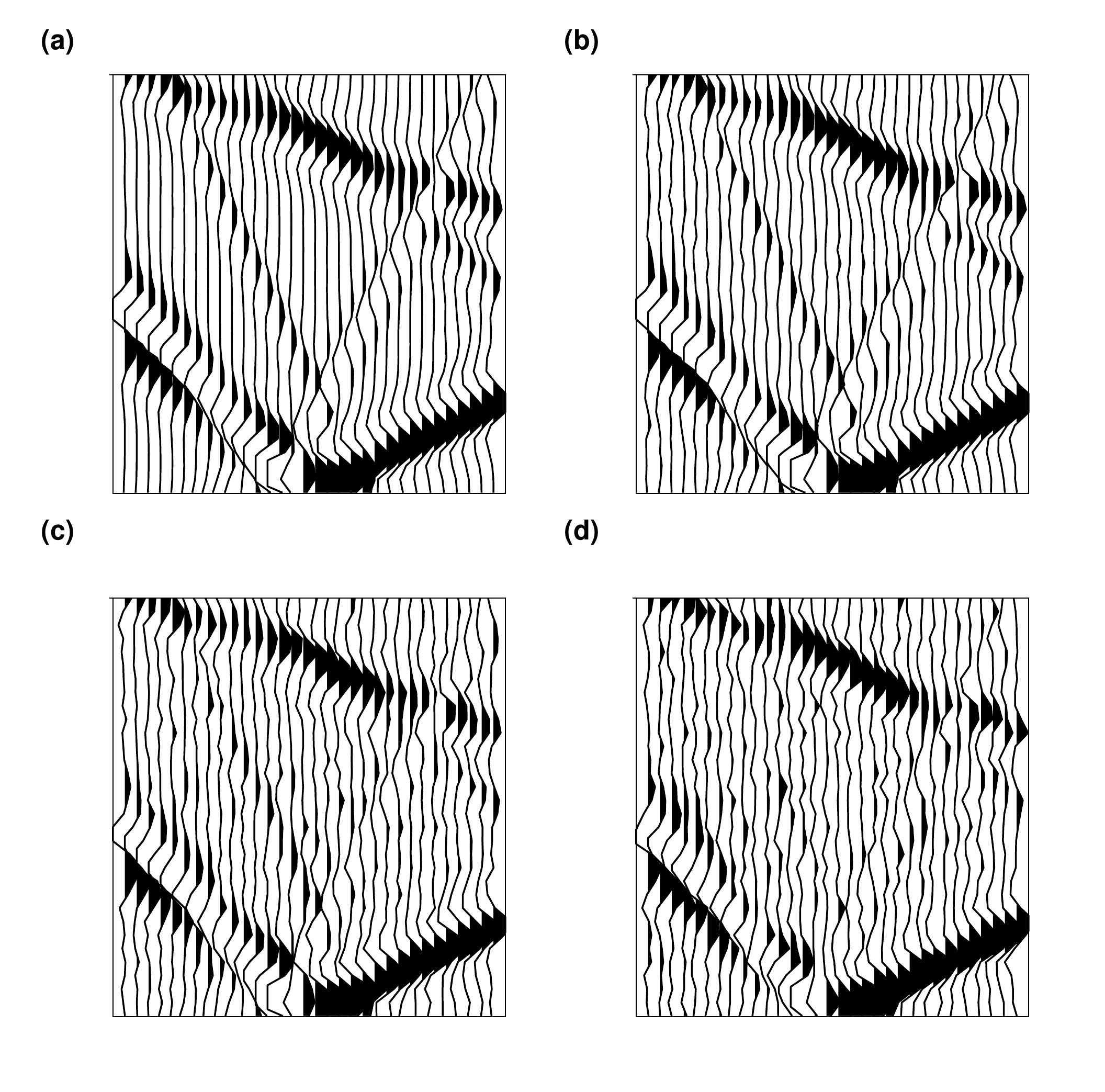}
   \vspace{-.75cm}  
   \caption{Performances of DnCNN and Deep-RED denoisers in compressive sensing recovery of seismic data. a) Clean data. b) Recovered data with $10\%$ compression. c) Recovered data with $25\%$ compression. d) Recovered data with $50\%$ compression. }
\label{fig4}%
\end{figure}

In this section, we evaluate the performances of Deep-RED regularizer in compressive sensing recovery of seismic data. To recover the signal of interest, similar to the previous section, we use the pre-trained DnCNN operators within the Deep-RED regularizer. Algorithm~\ref{alg2} is used to solve the optimization problem of Equation~\ref{n:12}. In Equation~\ref{n:12} we use sparse projection matrix
with a randomized discrete cosine transform~\cite{ailon2009fast} such that it transforms the signal from $\mathbb{R}^q$ to $\mathbb{R}^p$, where $\delta=\frac{p}{q}$ is the rate of compression. We apply the algorithm on clean data, similar to the de-noising section, with $32$ channels and $32$ time samples (Figure~\ref{fig4}a). In all of the synthetic examples, we use $\lambda=0.01$. 
Figure~\ref{fig4} shows the performances of the Deep-RED based recovery of compressive sensing measurements. To provide the measurements, we apply the compression matrix ${\bf A}_c$ on the clean signal ${\bf s}$, which is shown in Figure~\ref{fig4}a. We use $\delta={0.9, 0.75, 0.50}$, which results in $10\%, 25\%,$ and $50\%$ compressions, respectively. The un-compressed data recovered by using Algorithm~\ref{alg2} are shown in Figure~\ref{fig4}. Similar to synthetic examples in the de-noising section, here, we also report the highest quality of reconstruction. The quality of reconstruction on data with compression rates of $\delta={0.9, 0.75, 0.50}$, are $Q={15.16, 9.66, 6.92}$, respectively. The results show that the performance of the algorithm deteriorates as the compression rate decreases. Similar to the de-noising application, performances of the Deep-RED regularizer can be improved by either updating the training parameters of pre-trained DnCNN operators or by using algorithms that guarantee the Gaussianity of noise in each iteration. Nonetheless, in this paper, we only evaluate the across-domains transferability of Deep-RED, without modifying the training parameters. 
 \begin{figure*}[h]%
 \vspace{-0cm}
\centering  
   \includegraphics[width=0.95\textwidth]{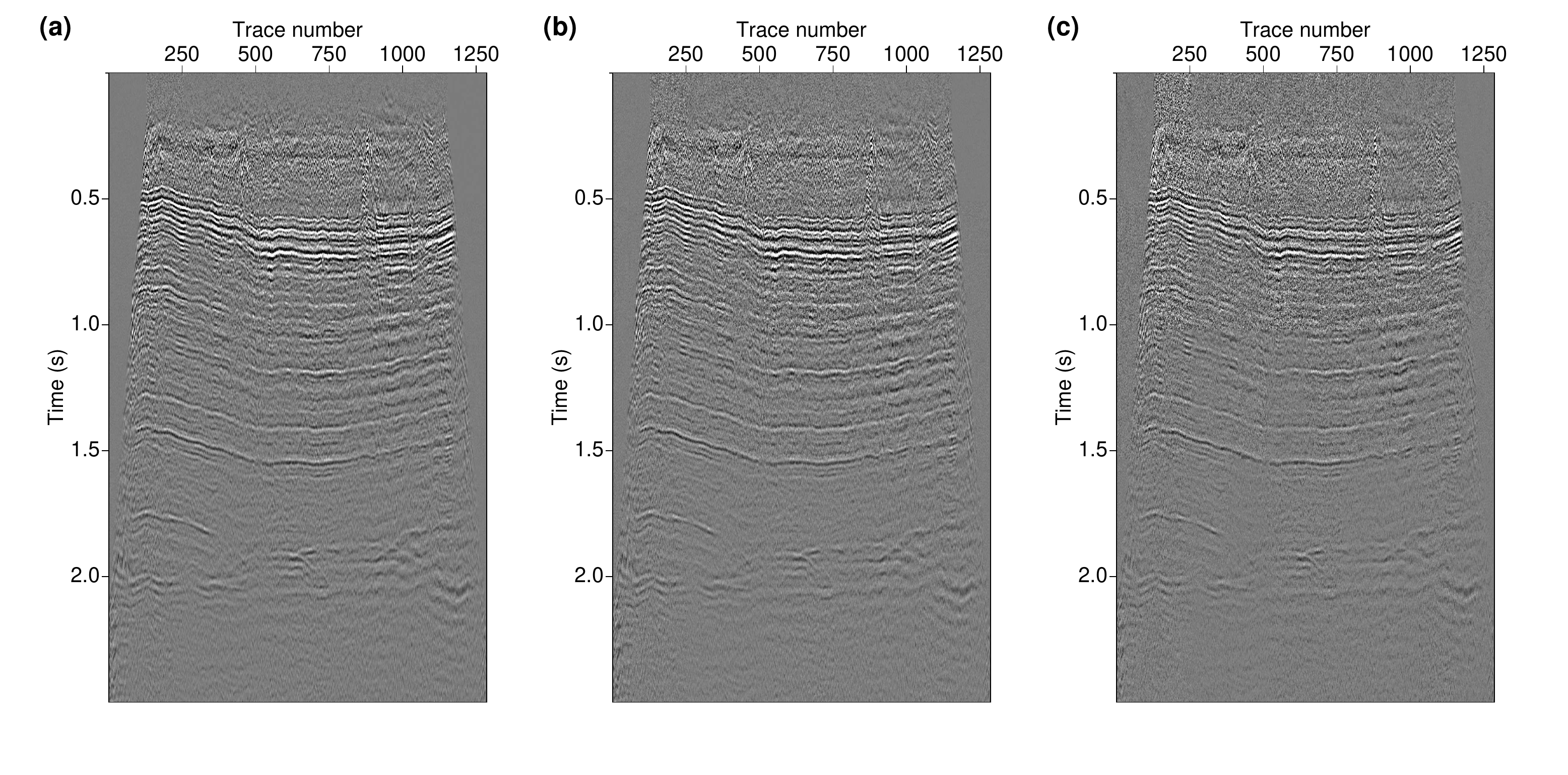}
   \vspace{-.75cm}  
   \caption{Performances of Deep-RED denoiser in compressive sensing recovery of real seismic data shown in Figure~\ref{fig3}a. a) Recovered data with $10\%$ compression. b) Recovered data with $25\%$ compression. c) Recovered data with $50\%$ compression. }
\label{fig5}%
\end{figure*}
We also test the Deep-RED based compressive sensing algorithm on the same real data that is used to demonstrate the de-noising performance of Deep-RED (Figure~\ref{fig3}a). In the case of compressive sensing recovery of real data, we do know the ground truth solution. However, in the realistic field measurements, we do not have access to ground-truth un-compressed signals as we only measure the compressed data. Hence, similar to the real data example in the de-noising section, we apply the DnCNN operators independently and report the solution that results in a maximum reduction in the cost function. All of the parameters, except the choice of DnCNN operator are kept the same. We use $\lambda=0.5$ and $\delta={0.9, 0.75, 0.50}$. The results are represented in Figure~\ref{fig5}. Results show that the Deep-RED is able to recover the seismic signal even though the DnCNN operators are trained on camera images only. However, the performance of the algorithm deteriorates as the rate of compression, $\delta$, decreases.            
\section{Conclusions}
We have developed an across-domains transferable Deep-RED denoiser for de-noising and compressive sensing recovery of seismic data. Through numerical and real-world data examples, we have shown that it is possible to use the DnCNN operators, which are merely trained on camera images, to de-noise and recover the compressively sensed seismic data. The key feature for the success of such applications was that the signal processing algorithms implement similar formula and optimization methods. Accordingly, by incorporating the DnCNN operator in RED regularizer and using well-known signal processing optimization methods, we were able to successfully transfer the DnCNN operator from camera image processing to seismic data processing domains. We have tested the proposed algorithms on de-noising and compressive sensing recovery of seismic data, without updating the training parameters of the pre-trained DnCNN operators. In de-noising examples, we have successfully de-noised synthetic and real seismic data. The performance of the algorithm, however, deteriorated when the SNR of data is decreased. Later, we have shown the performance of the Deep-RED in compressive sensing recovery of synthetic and real seismic data. The results showed that the method recovers the signal on data with a compression rate as low as $\delta=0.5$, efficiently. 
\bibliographystyle{IEEEtran}
\bibliography{Biblio}

\end{document}